\documentclass[a4paper,10pt]{scrartcl}

\usepackage{amsmath}
\usepackage{amsfonts}
\usepackage{amssymb}
\usepackage{graphicx}
\usepackage[english]{babel}
\usepackage{wasysym}
\usepackage{units}

\newcommand{\omegaK}{\omega_{\textrm{\tiny K}}}

\newcommand{\reff}[1]{(\ref{#1})}

\newcommand{\qq}{\qquad}
\newcommand{\q}{\quad}
\newcommand{\vs}[1]{\vspace{#1mm}}
\newcommand{\vsO}{\vspace{.1cm}\hfill\\}
\newcommand{\vsT}{\vspace{.2cm}\hfill\\}

\title{\large EMERGENCE OF HIGH PEAKS IN THE AXIAL VELOCITY FOR\\
AN IDEAL MAGNETOHYDRODYNAMICS DISK CONFIGURATION}

\author{\normalsize Giovanni Montani$^{\;a,\;b,\;c}$ and Nakia Carlevaro$^{\;a}$\vsT
\emph{\footnotesize $^a$ Department of Physics - ``Sapienza'' University of Rome}\vs{-2.5}\\
\emph{\footnotesize c/o Dip. Fisica - ``Sapienza'' Universit\`a di Roma, P.le A. Moro, 5 (00185), Roma (Italia).}\vs{-2}\\
\emph{\footnotesize $^b$ 
ENEA – C.R. Frascati (Rome), UTFU-MAG.}\vs{-2}\\
\emph{\footnotesize $^c$
ICRANet -- International Center for Relativistic Astrophysics Network.}\\
{\footnotesize\ttfamily giovanni.montani@frascati.enea.it\quad\qquad nakia.carlevaro@gmail.com}
}
\date{}
\begin{document}
\maketitle

%
\hrule
\begin{abstract} \textbf{Abstract:} We study the profile of a thin disk configuration as described by an axisymmetric ideal magnetohydrodynamics steady equilibrium. We consider the disk like a differentially rotating system dominated by the Keplerian term, but allowing for a non-zero radial and vertical matter flux. As a result, the steady state allows for the existence of local peaks for the vertical velocity of the plasma particles, though it prevents the radial matter accretion rate. This ideal magnetohydrodynamics scheme is therefore unable to solve the angular momentum-transport problem, but we suggest that it provides a mechanism for the generation of matter-jet seeds.

\vsO \emph{PACS}: 97.10.Gz; 95.30.Qd; 52.30.Cv
\end{abstract}
\hrule

\paragraph{1. Basic Statements -} In this paper, we consider an axisymmetric ideal magnetohydrodynamics (MHD) model for the disk structure surrounding a compact astrophysical object. Our approach, which differs from the standard astrophysical model for the disk morphology, is justified in view of the microphysical properties of the plasma having negligible value of the viscosity and the resistivity coefficients. Indeed, the standard model \cite{S73, B01} accounts for the angular momentum transport by postulating a non-zero viscous stress tensor appearing in the equilibrium as an effect of the internal turbulence. Despite this model is rather successful in explaining the observations, it remains affected by the non trivial shortcoming that the microscopic estimations for the viscosity coefficient give much smaller values than those ones required to fit data. This is particularly true in accreting objects like X-ray stars, where the observed rate of accretion is very high. An analogous difficulty concerns the proper balance of the force acting on electrons, when a significant infalling radial velocity occurs. The solution to this puzzle is offered by the presence of an effective resistivity of the plasma, which compensate the azimuthal Lorentz force acting on the electrons. Also such values of the resistivity coefficient are in contradiction with the microscopical estimations related to the astrophysical scales. The turbulent viscosity and resistivity coefficients can be maintained both relevant for the equilibrium by postulating a \emph{Prandtl number} of order unity for the accreting plasma. This shortcoming of the standard model is still open to the scientific debate and leads to speak of an \emph{anomalous} visco-resistivity of the disk plasma, rising from the so-called magnetorotational instability described by E. Velikhov \cite{V59, B98}. However, the laboratory experience on turbulent plasmas does not confirm the correlation between the internal instabilities and the applicability of a viscoresistive MHD model with Prandtl numbers of order unity.\\
$\phantom{11}$In our analysis, we address a different scenario as closely related to the transport profile of magnetically confined laboratory plasmas \cite{C94}. Indeed, in \cite{O97}, has been shown how the ideal MHD profile of a thin disk can have a significant deviation from the Keplerian behavior for a strong vertical Lorentz force. In \cite{C05, CR06} it has been demonstrated that the local configuration of a thin disk can be reduced to the profile of a ring sequence, when the electromagnetic backreaction is sufficiently intense and the vertical Lorentz force is able to confine the disk (see also \cite{C08}). Despite a solution of the angular momentum transport (\emph{i.e.}, a non-zero radial accretion rate) is not jet available in this ideal scheme, it suggests the idea that (instead of a turbulent behavior) the disk plasma could be characterized by the formation of microstructures which might drive the typical \emph{ballooning-mode} instabilities observed in laboratory systems. The accretion mechanism would then rely on the porosity of the plasmas near the x-points of the magnetic field configuration.\\
$\phantom{11}$We here follow the research line traced in \cite{O97, C05, CR06}, but including vertical and radial matter fluxes in the ideal MHD analysis (see also \cite{CR07, MB09, LM10}). The resulting equilibrium configuration is characterized by a vanishing net mass accretion rate, as in the previous approaches, since contributions due to the radial velocity over and below the equatorial plane cancel out. However, our local configuration permits the existence of specific disk layers where the vertical velocity can acquire a very large amplitude. This scenario suggests the possibility to realize matter jets from the disk equilibrium profile, especially in view of the implementation of this same morphology in presence of the expected plasma instabilities and this issue constitutes the main achievement of our analysis. Indeed, the link between highly magnetized disk plasmas and the jet formation (observed in the different electromagnetic bands) constitutes, in the standard model, one of the most relevant motivations for studying the angular momentum transport across these axisymmetric systems (for a proposal on the jet formation see \cite{LB96}). Our result appears particularly appealing because the viscoresistive model has some difficulties to create the conditions for the jet formation, because the viscoresistive MHD scenario predicts diffusive magnetic configurations \cite{S08, SS01}.

\paragraph{2. Vertical and Radial Equilibria -} To adapt the MHD equations to the axial symmetry, we introduce cylindrical coordinates $\{r,\;\phi,\;z\}$ and the fundamental variables have no dependence on $\phi$. The gravitational potential of the central object writes
\begin{equation}
\chi(r,z)=GM_\mathrm{S}\;/\;\sqrt{r^2+z^2}\;,
\label{Gravpot}
\end{equation}
where $M_S$ is the star mass and $G$ the Newton constant. The continuity equation, in the steady state of the disk, reads as the vanishing divergence of the matter flux $\epsilon \vec{v}$ ($\epsilon$ being the mass density and $\vec{v}$ the velocity field) and it admits the following solution:
\begin{equation}
\epsilon \vec{v}= 
-\vec{e}_r\;(\partial_z\Theta)/r +
\vec{e}_{\phi}\;\epsilon r\,\omega(r,z^2)+
\vec{e}_z\;(\partial_r\Theta)/r\;,
\label{solconteq}
\end{equation}
where, $\omega (r,z^2)$ is the disk differential angular velocity (in view of the co-rotation theorem \cite{F37}, it can be written as $\omega = \omega(\psi)$, $\psi(r,z^2)$ being the magnetic flux surface). Here, the generic function $\Theta(r,z)$ must be odd in $z$ to deal with a non-zero mass accretion rate of the disk ($\dot{M}_\mathrm{d}$):
\begin{equation}
\dot{M}_\mathrm{d} = -2\pi r\int _{-H_0}^{+H_0}dz\;\epsilon v_r =
4\pi\; \Theta (r,H)\;>\;0\;,
\label{Mdot}
\end{equation}
where $H_0$ is the half-depth of the disk and $v_r<0$ the infalling radial velocity. The magnetic field reads as
\begin{equation}
\vec{B} = -\vec{e}_r\;(\partial_z\psi)/r +
\vec{e}_\phi\;I(\psi,z)/r +
\vec{e}_z\;(\partial_r\psi)/r\;,
\label{vectorb}
\end{equation}
with the function $I(\psi,z)$ allowing for the existence of poloidal currents in the disk plasma.

In order to investigate the effects induced on the disk profile by the electromagnetic reaction of the plasma, we consider a local model of the equilibrium around a radius value $r=r_0$. Thus, we split the mass density $\epsilon$ and the thermostatic pressure $p$ into the background (barred terms) and perturbation (hatted terms) components, as $\epsilon=\bar{\epsilon}(r_0,z^2)+\hat{\epsilon}$ and, of course, $p=\bar{p}(r_0,z^2)+\hat{p}$, respectively. The same way, we express the magnetic surface function in the form $\psi=\psi_0(r_0)+\psi_1(r_0,r-r_0,z^2)$, with $\psi_1\ll\psi_0$. In particular, the co-rotation theorem allows the representation $\omega=\omegaK +\tilde{\omega}_0\psi_1$, where $\omegaK$ is the Keplerian term and $\tilde{\omega}_0=const$. This form for $\omega$ holds locally, as far as $(r-r_0)$ remains sufficiently small so that the dominant deviation from the Keplerian contribution is due to $\psi_1$.

When we consider the ``drift ordering'' for the behavior of the gradient amplitude (\emph{i.e.}, the first-order gradients of the perturbations are of zeroth-order while the second-order ones dominate) we are able to cast the equilibrium configuration by a hierarchical ordering of the different contributions. Accordingly, the profile of the disk toroidal currents $J_{\phi}$ and the azimuthal component of the Lorentz force $F_{\phi}$ have the following form:
\begin{subequations}
\begin{align}
J_{\phi}&\simeq -c\;(\partial^2_r\psi_1 + \partial_z^2\psi_1)/4\pi r_0\;,
\label{jphi}\\
F_{\phi}&\simeq (\partial_zI\;\partial_r\psi-\partial_r\;I\partial_z\psi)/4\pi r_0\;.
\label{fphi}
\end{align}
\end{subequations}

We now fix the equations governing the vertical and the radial equilibria of the disk by separating the fluid components from the electromagnetic backreaction (as discussed in \cite{C05, CR06}). Indeed, we assume that the contribution of the toroidal magnetic field does not affect these equilibria and therefore the resulting configuration equations overlap that ones derived originally by B. Coppi in 2005. The quantity $I(\psi,z)$ plays a crucial role in the azimuthal equation only, allowing for a non-zero radial velocity even in the ideal MHD scenario. 

The splitting of the MHD equations gives, for the vertical force balance, the following system of equations
\begin{subequations}
\begin{align}
&D\equiv\bar{\epsilon}/\epsilon_0=\exp[-z^2/H_0^2]\;,\;\;\;\;
H_0^2\equiv 4K_\mathrm{B}T/(m\omegaK^2),\\
&\partial_z\hat{p}+\omegaK^{2} z\hat{\epsilon}-
\partial_z\psi_1\left(\partial^2_z\psi_1+\partial^2_r\psi_1\right)/4\pi r_0^2=0\;,
\label{verticalequilibrium}
\end{align}
\end{subequations}
where $\epsilon_0=\epsilon_0(r_0)\equiv\epsilon(r_0,0)$ and the temperature $T$ admits the form ($K_\mathrm{B}$ being the Boltzmann constant):
\begin{equation}
2K_\mathrm{B}T\equiv m(p/\epsilon)=
m(\bar{p}+\hat{p})/(\bar{\epsilon}+\hat{\epsilon})\equiv 2 K_\mathrm{B}(\bar{T}+\hat{T})\;,
\label{temptot}
\end{equation}
The radial equilibrium splits into the decomposition for $\omega$ (\emph{i.e.}, $\omega\simeq\omegaK+\delta\omega\simeq\omega_0(\psi_0)+\tilde{\omega}_{0}\psi_1$) and the equation
\begin{align}
&2\omegaK r_0(\bar{\epsilon}+\hat{\epsilon})\tilde{\omega}_0\psi_1
+\partial_r\psi_1\,(\partial^2_z\psi_1+\partial^2_r\psi_1)/(4\pi r_0^2)=\nonumber\\
&\qq\qq=\;\partial_r[\hat{p}+(\partial_r\psi_1)^2/(8\pi r_0^2)]+
\partial_r\psi_1\partial^2_z\psi_1/(4\pi r_0^2)\;.
\label{radialequilibrium}
\end{align}

Let us now define the dimensionless functions $Y$, $\hat{D}$ and $\hat{P}$, in place of $\psi_1$, $\hat{\epsilon}$ and $\hat{p}$, \emph{i.e.},
\begin{equation}
Y\equiv k_0\psi_1/(\partial_{r_0}\psi_0)\;,\quad
\hat{D}\equiv \beta\hat{\epsilon}/\epsilon_0\;,\quad
\hat{P}\equiv \beta\hat{p}/p_0\;,
\label{deff}
\end{equation}
where $p_0\equiv 2K_\mathrm{B}\bar{T}_0\epsilon _0$ ($\bar{T}_0$ being $\bar{T}(r,0)$) and 
\begin{align}
\beta \equiv 8\pi p_0/B^2_{0z}=1/(3\epsilon _z^2)\equiv k_0^2H_0^2/3\;.
\end{align}
Here, we introduced the fundamental wave-number $k_0$ of the radial equilibrium, defined as $k_0\equiv 3\omegaK^2/v_\mathrm{A}^2$, with $v_\mathrm{A}^2\equiv B^2_{z0}/(4\pi\epsilon_0)$, recalling that $B_{z0}=(\partial_{r_0}\psi_0)/r_0$. Thus, we introduce the dimensionless radial variable defined as $x\equiv k_0(r-r_0)$, while the fundamental length in the vertical direction is taken as $\Delta\equiv\sqrt{\epsilon_z}H_0$, leading to define $u\equiv z/\Delta$. This way, the relation $D=\exp[-u^{2}\epsilon_z]$ holds.

These definitions allow us to rewrite the vertical and radial equilibria as follows
\begin{subequations}\label{equlibbbbbbb}
\begin{align}
&\partial_{u^2}\hat{P} + \epsilon_z\hat{D}-
2\left(\partial^2_{x}Y + \epsilon_z\partial^2_{u}Y\right)\partial_{u^2}Y=0\;,\label{vertad}\\
&(D+\hat{D}/\beta)Y+\partial^2_{x}Y+\epsilon_z\partial ^2_{u}Y+\nonumber\\
&\qq\qq+\tfrac{1}{2}\,\partial_x\hat{P}+\left(\partial^2_{x}Y+\epsilon_z\partial^2_{u}Y\right)
\partial_x Y=0\;.\label{radad}
\end{align}
\end{subequations}
The system above fixes two of the three unknowns $Y$, $\hat{P}$ and $\hat{D}$ and has to be completed by the azimuthal equilibrium and the electron force balance. We stress that, in setting eqs.\reff{equlibbbbbbb}, we have neglect the inertial contributions coming from the radial and the vertical components of the matter velocity, because of their small amplitude. 

\paragraph{3. Additional Configuration Equations -} The equilibrium along the toroidal symmetry of the disk writes
\begin{equation}
\epsilon v_r\partial _r(\omega r) +
\epsilon v_z\partial _z(\omega r) +
\epsilon \omega v_r = F_{\phi }\;.
\label{exazeq}
\end{equation}
Using the co-rotation theorem, \emph{i.e.}, $\omega=\omega(\psi)$, and eqs.\reff{vectorb}, \reff{fphi}, the equation above restates as
\begin{align}
\epsilon r_0^2( & v_r B_z - v_z B_r) + 2\epsilon v_r\omegaK/ \tilde{\omega}_0
=\nonumber\\
&=\left[\partial _zI\left( \partial _{r_0}\psi_0
+ \partial _r\psi _1\right) 
 - \partial_rI\partial _z\psi_1\right]/(4\pi r_0^2\tilde{\omega}_0)\;.
\label{exazeqlo}
\end{align}
In this scheme, we link the radial and vertical velocity fields to the poloidal currents. Indeed, we aim to set up a picture in which the radial matter flux is induced by the poloidal currents, \emph{i.e.}, establishing a direct relation between $v_r$ and $F_{\phi}$. To this end, as well as for the consistence of the model, we have to analyze the electron force balance equation: $\vec{E}+(\vec{v}/c)\wedge\vec{B}=0$. 
The balance of the Lorentz force has radial and vertical components which express the electric field in the form predicted by the co-rotation theorem \cite{F37, C05} and, since the axial symmetry requires $E_{\phi}\equiv0$, the $\phi$-component of the this equation retains the form
\begin{equation}
v_z B_r - v_r B_z = 0\;.
\label{efb1x}
\end{equation}
We now observe that, substituting eq.(\ref{efb1x}) in eq.(\ref{exazeqlo}), we get the searched relation between $v_r$ and $F_{\phi}$, \emph{i.e.},
\begin{align}
2\epsilon v_r \omegaK =[\partial _zI\left( \partial _{r_0}\psi_0
+ \partial _r\psi _1\right)-\partial_rI\partial _z\psi_1]/(4\pi r_0^2)\,.
\label{exazeqlocomb}
\end{align}
This relation, together with eq.(\ref{efb1x}), provides a system for the two unknown functions $\Theta$ and $I$, once $\psi_1$ and $\epsilon$ are given by the vertical and the radial equilibria, \emph{i.e}, eq.(\ref{vertad}) and eq.(\ref{radad}), respectively.

Substituting $v_r$ expressed by eq.(\ref{exazeqlocomb}) into the electron force balance (\ref{efb1x}), we get the following expression for $v_z$:
\begin{align}\label{efblocn}
v_z =- \frac{\left( \partial _{r_0}\psi _0 + \partial _r\psi _1\right) 
	F_{\phi }}{2\omegaK \epsilon _0\partial_z\psi_1(D+\hat{D}/\beta)}\;.
\end{align}
We stress that, addressing the splitting $\psi=\psi_0+\psi_1$, the azimuthal Lorentz force stands as
\begin{equation}
F_{\phi} \simeq \left[\partial _zI\left( \partial _{r_0}\psi_0
+ \partial _r\psi _1\right) 
- \partial_rI\partial _z\psi_1\right]/(4\pi r_0^2)\;.
\label{f0hispl}
\end{equation}
Expressing eq.(\ref{exazeqlocomb}) for $v_r$ and eq.(\ref{efblocn}) for $v_z$ via the function $\Theta$, we get the following integro-differential compatibility condition
\newcommand{\intt}{{\textstyle{\int}}}
\begin{equation}
\partial_z\psi_1\;[\intt dz\;\partial_r F_{\phi}]=F_{\phi}
\left(\partial_{r_0}\psi_0 + \partial_r\psi_1\right)\;,
\label{fphicompdim}
\end{equation}
from which, provided $\psi_1$, \emph{i.e.}, $Y(x,u)$, from eqs.\reff{equlibbbbbbb}, one can determine the toroidal magnetic field $I(\psi,z)$.

\paragraph{4. Non-Linear Limit -} Let us now analyze the case $Y\gg1$, \emph{i.e.}, when the electromagnetic reaction within the disk plasma is strong enough to significantly deform the background magnetic field. In this context, the vertical and radial equilibria are now described by eq.(\ref{vertad}) and
\begin{align}
(D+\hat{D}/\beta)Y +\tfrac{1}{2}\partial_x\hat{P}+
(\partial^2_x Y+\epsilon_z\partial^2_u Y)\partial_x Y = 0\,,\label{radadnl}
\end{align}
respectively. Clearly, we got a new configuration system for which linear terms in the perturbed magnetic surface $Y$ have been neglected in favor of the quadratic ones. In the same limit, the azimuthal equation (\ref{exazeqlocomb}) rewrites
\begin{equation}
2\epsilon v_r \omega _K = F_{\phi} = 
\left[\partial _zI\partial _r\psi _1          
 - \partial_rI\partial _z\psi_1\right]/(4\pi r_0^2)\;,
\label{exazeqlocombnl}
\end{equation}
and eqs.(\ref{efblocn}), \reff{fphicompdim} rewritten in terms of $Y$ as follows:
\begin{subequations}
\begin{align}
&v_z=-[v_r\,\partial_x Y]/[\sqrt{\epsilon_z}\;\partial_u Y]\;,\label{efbnl}\\
&\partial_uY\;[\intt du\;\partial_x F_\phi]=F_{\phi}\partial_x Y \;,\label{fphicompdimnl}
\end{align}
\end{subequations}
respectively. We now look for a separable solution of the configurational system of the form
\begin{subequations}\label{solarray}
\begin{align}
Y(u^2,x) &= F(u^2)\sin (\alpha x)\;,\\
\hat{D}(u^2,x)&= d(u^2)\cos (\alpha x)\;,\\ 
\hat{P}(u^2,x) &= \xi(u^2)\sin ^2(\alpha x) + \pi (u^2)\cos (\alpha x)\;,\\
F_{\phi}(u^2,x) &= G(u^2)\sin^b(\alpha x)\;.
\end{align}
\end{subequations}
Substituting eqs.\reff{solarray} in the set of equations derived above, we get ordinary differential relations in the variable $u$: from eqs.(\ref{vertad}), (\ref{radadnl}), we get the following system
\begin{subequations}\label{verordsy}
\begin{align}
\partial_{u^2}\pi+\epsilon_z d&=0\;,\\
\pi-2DF/\alpha&=0\;,\\
\partial_{u^2}\xi+2\left(-\alpha^2 F+\epsilon_z F^{\prime\prime}\right)\partial_{u^2}F&=0\;,\\
3\epsilon_z^2Fd+\alpha\xi+\alpha F(-\alpha^2 F+\epsilon_z F^{\prime\prime})&=0\;.
\end{align}
\end{subequations}
Using $\epsilon=\bar{\epsilon}+\hat{\epsilon}$, eq.(\ref{exazeqlocombnl}) and eq.\reff{efbnl} give
\begin{subequations}
\begin{align}
v_r&=[G\sin^b(\alpha x)]\;/\;[2\omega_K\epsilon_0(D+d\cos(\alpha x)/\beta)]\;,\label{aziordsy}\\
v_z&=-[\alpha F\;\textrm{cotg}(\alpha x)\; v_r]\;/\;[F^{\prime}\sqrt{\epsilon _z}]\;,
\end{align}
\end{subequations}
respectively. Finally, the compatibility equation \reff{fphicompdimnl} can be recast to give the integro-differential relation:
\begin{equation}
bF^{\prime}\;[\intt du\;G]=FG\qquad\Rightarrow\qquad
[\intt du\; G] = gF^b\;,
\label{fphicompdim2}
\end{equation}
where we have used $(...)^{\prime}=d(...)/du$ and $g=const$.

\begin{figure}[ht]
\centering
\includegraphics[width=.47\textwidth,clip]{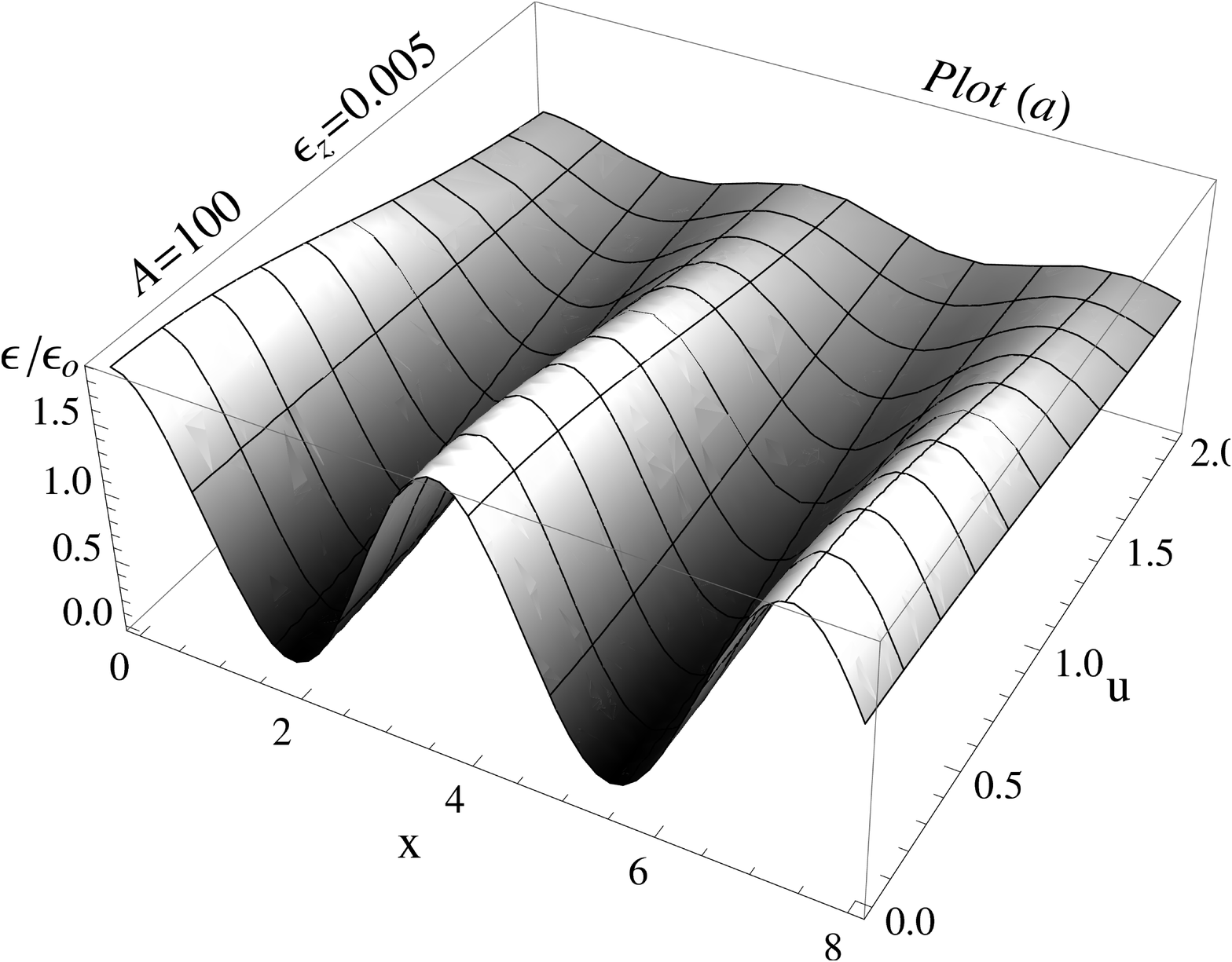}
\includegraphics[width=.47\textwidth,clip]{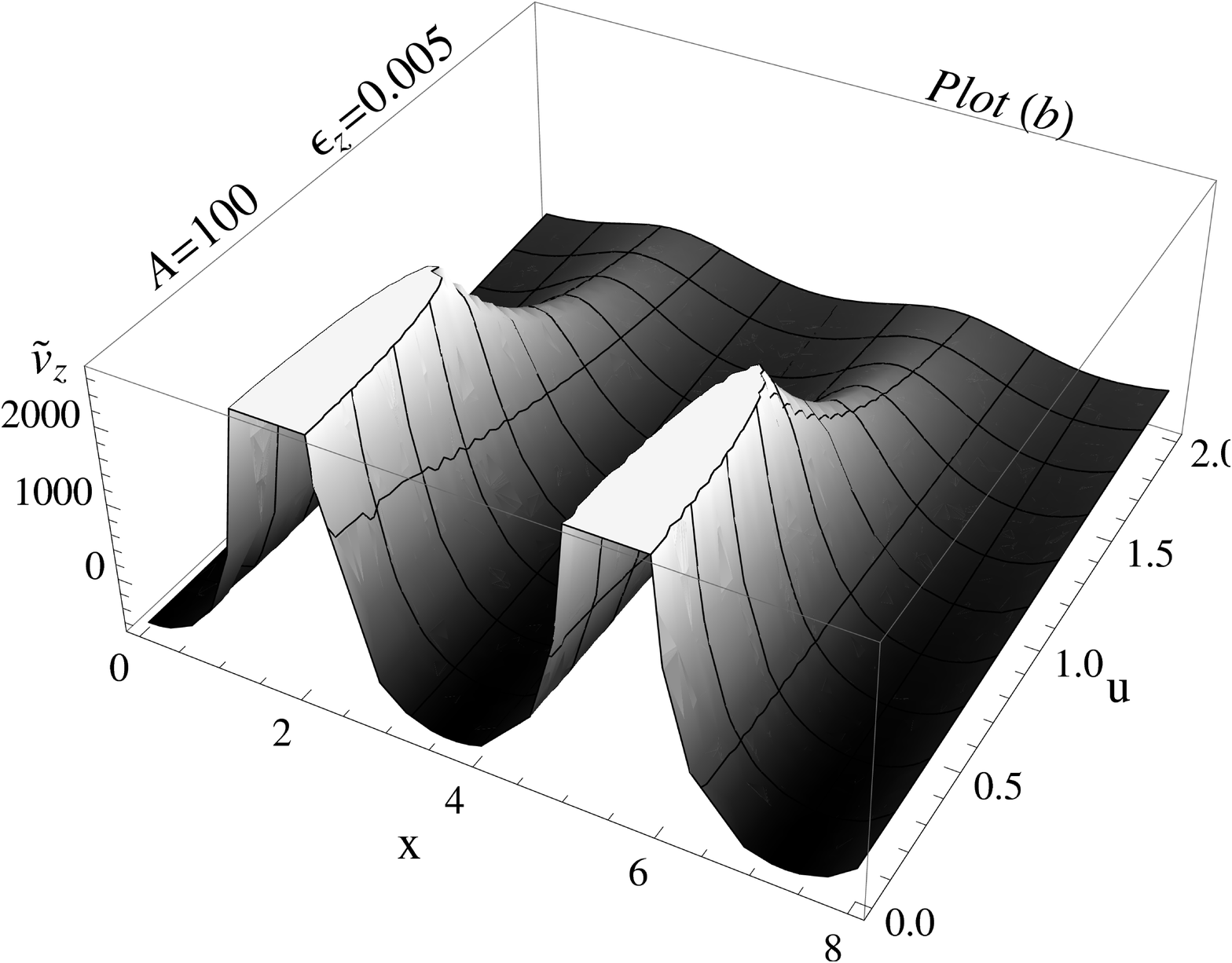}
\caption{Behavior of $\epsilon/\epsilon_0$ (Plot (a)) and $\tilde{v}_z=v_z/v^{*}$ (Plot (b)) ($v^{*}=g/\omegaK\epsilon_0$), setting $A=100$ and $\epsilon_z=0.005$.}
\label{fig:plot-density}
\end{figure}

\paragraph{5. Analytical Solution -} To derive an exact solution of the system fixed above, we now consider the case of large $\beta$-values for the disk plasma, \emph{i.e.}, $\epsilon _z\ll 1$. Such analytical solution writes
\begin{subequations}\label{exsol}
\begin{align}
D&=1-\epsilon_z u^2\;,\q
F=A\exp[-u^2/2]\;,\q
A\gg1\;,\\
\alpha &=\sqrt{3}\;,\q
d=F/(\epsilon_z\sqrt{3})\;,\q
\xi=(3-\epsilon_z u^2)F^2\;,\\
\pi&=2(1-\epsilon_z u^2)F/\sqrt{3}\;,\qq
G=-bguF^b\;.
\end{align}
\end{subequations}
Hence the radial velocity \reff{aziordsy} takes the explicit form
\begin{equation}
v_r=-\frac{1}{2\omega_K\epsilon_0}\;
\frac{bgA^b u e^{-bu^2/2}\sin^b(\sqrt{3}x)}{
1+\sqrt{3}\,\epsilon_z A e^{-u^2/2} \cos(\sqrt{3}x)}\;,
\label{aziordsya}
\end{equation}
where, here and in the following, we neglect the term $\epsilon_z u^2$ with respect to terms proportional to $A$. It is remarkable that the total mass density is expressed by the relation $\epsilon=\epsilon_0(D+\hat{D}/\beta)$, \emph{i.e.},
\begin{equation}
\epsilon=\epsilon_0\big[1+\sqrt{3}\,\epsilon_z A e^{-u^2/2}
\cos(\sqrt{3}x)\big]\;,
\label{eeeeeeeee}
\end{equation}
and $\epsilon\geqslant0$ implies $\sqrt{3}\epsilon_zA \leqslant 1$. This feature of the mass density profile (Figure \ref{fig:plot-density}, Plot (a)) 
ensures that the ring sequence, predicted in \cite{CR06}, is present in this regime too. Since $v_r$ is odd in $u$ (\emph{i.e.}, $\Theta$ is even in $z$) the accretion rate of the disk vanishes identically. Despite a local non-zero radial matter flux is present in the equilibrium configuration, however the compensation between the contributions over and below the equatorial plane, respectively, gives a net zero accretion of the disk.

Finally, the vertical velocity component reads as
\begin{equation}
v_z=-\frac{\sqrt{3/\epsilon_z}}{2\omega_K\epsilon_0}\;
\frac{b g A^b e^{-bu^2/2} \sin^{b-1}(\sqrt{3}x) \cos(\sqrt{3}x)}{
1+\sqrt{3}\,\epsilon_z A e^{-u^2/2} \cos(\sqrt{3}x)},
\label{aziverdsya}
\end{equation}
Let now consider the peculiar case $b=1$ of eq.\reff{aziverdsya}, near the equatorial plane. For $x\sqrt{3}\sim\pi$, we deal with a positive or negative vertical velocity (depending on the sign of $Ag$) having a very large amplitude, $\mid v_z\mid\gg1$ (Figure \ref{fig:plot-density}, Plot (b)), as far as the condition $\epsilon_z\sqrt{3}\,A\sim1$ holds (\emph{i.e.}, in correspondence of the mass-density nodes). Such a condition is satisfied only in specific layers of the disk configuration since our model is localized at $r_0$ but $A$, $g$ and $\epsilon_z$ vary with it. 

Finally, the total term describing the magnetic surface, once we account for the first-order corrections around $r_0$, reads
\begin{equation}
Y_t\equiv x+Y=x+A\exp[-u^2/2]\sin(\sqrt{3}x)\;.
\label{tms}
\end{equation}
The ratios of the magnetic field components to the background magnetic field due to the central object, \emph{i.e.}, $\vec{B}_0 = {B_z}_0\vec{e}_z$, read as 
\begin{subequations}\nonumber
\begin{align}
B_z/B_{z0}&=\partial_x Y_t\simeq\,\sqrt{3}\,A\exp[-u^2/2]\cos(\sqrt{3}x)\;,\\
B_r/B_{z0}&=-\sqrt{\epsilon_z}\partial_u Y_t=\,\sqrt{\epsilon _z}\,Au\exp[-u^2/2]\sin(\sqrt{3}x)\;.
\end{align}
\end{subequations}

We stress that the condition to neglect the inertial terms due to the radial and vertical velocity (despite of the presence of $v_z$-peaks) in the system \reff{equlibbbbbbb} takes the following form: $Ag/(1-\sqrt{3}\epsilon_z A)\sim\mathcal{O}(1)$. Moreover, in the regime $\epsilon_z\ll1$ (for which our solution is derived), the condition to deal with a positive mass density, \emph{i.e.}, $\sqrt{3}\epsilon_zA \leqslant 1$, ensures the positivity of the total pressure too.

\paragraph{6. Astrophysical Characterization -} The configuration scheme we adopted above is appropriate to describe a local radial layer (a small circular crown) of a thin accretion disk since all the morphological prescriptions of this astrophysical system are included. The choice of the background magnetic surface as a function of $r_0$ only, models a dipole magnetic-field of the central object, as far as the disk is sufficiently thin and we take $\omegaK\propto\psi_0^{3}$. The gravitational field is correctly represented in the thin disk approximation, for the limit of an high Keplerian angular velocity to have a gravostatic confinement of the vertical profile of the disk, \emph{i.e.}, $\omegaK^2 z\sim \mathcal{O}(1)$. The idea that the poloidal (radial and vertical) matter fluxes are taken small with respect to the toroidal Keplerian motion is a natural approximation, valid in the standard model too. The vertical velocity must be smaller than the azimuthal one in a thin disk, because of the background vertical equilibrium, while the radial motion is required perturbative since its presence must preserve the steady character of the matter distribution. In the standard model the amplitude of $v_r$ is determined by the turbulent viscosity coefficient, while here it is fixed via the self-consistency of the electron force balance equation. Indeed, we deal with a vanishing net accretion, but, from a physical point of view, we could speak of a weak constant rate of matter infall, since this fact do not affect our main goal about the vertical matter jets. In this ideal plasma scenario, we can reach very high self-consistent magnetic fields and we show how they can induce intense axial matter fluxes. This picture is commonly believed the basic mechanism of the emission from high energy astrophysical sources (such as Gamma Ray Bursts or Active Galactic Nuclei), but its setting is still rather questioned. We offer here an alternative explanation, surprisingly relying only on the morphology of ideal and highly non-linear MHD structures.

\paragraph{7. Conclusions -} We have shown how the steady state of an accretion disk, described by ideal MHD, is compatible with the presence of significant radial and vertical matter fluxes. Such matter currents are unable to account for a non-zero accretion rate, but they are relevant in view of the instabilities that can be driven by their existence. In particular, we outlined, as the main issue of our study, that there are conditions under which strong vertical matter fluxes are locally allowed, especially near the equatorial plane. This feature deserves attention because dealing with matter jets along the symmetry axis is one of the expected features of the disk configurations, but it is hard to be recovered in the standard viscoresistive model.\\
{\footnotesize $^{**}$ This work was developed within the framework of the \emph{CGW Collaboration} (www.cgwcollaboration.it).$^{**}$}

{\small\textbf{Acknowledgment}: NC gratefully acknowledges the CPT - Universit\'e de la Mediterran\'ee Aix-Marseille 2 and the financial support from ``Sapienza'' University of Rome.}


\newpage
\footnotesize
\begin{thebibliography}{0}

\bibitem{S73}
N.I. Shakura, \emph{Sov. Astron.} \textbf{16}, 756 (1973).

\bibitem{B01}
G.S. Bisnovatyi-Kogan, R.V.E Lovelace, \emph{New. Astron. Rev.} \textbf{45}, 663 (2001).

\bibitem{V59}
E. Velikhov, \emph{Sov. Phys. JETP} \textbf{36}, 995 (1959).

\bibitem{B98}
S.A. Balbus, J.F. Hawley, \emph{Rev. Mod. Phys.} \textbf{70}, 1 (1998).

\bibitem{C94}
B. Coppi, \emph{Plasma Phys. Contrl. Fus.} \textbf{36}, B107 (1994).

\bibitem{O97}
G.I. Ogivile, \emph{Mon. Not. RAS} \textbf{288}, 63 (1997).

\bibitem{C05}
B. Coppi, \emph{Phys. Plasmas} \textbf{12}, 057302 (2005).

\bibitem{CR06}
B. Coppi, F. Rousseau, \emph{Astrophys. J.} \textbf{641}, 458 (2006).

\bibitem{C08}
B. Coppi, \emph{EuroPhys. L.} \textbf{82}, 19001 (2008).

\bibitem{CR07}
B. Coppi, F. Rousseau, in the proceedings of \emph{The 34$^{th}$ EPS Conference on Plasma Physics}, Warsaw (July 2007), Paper O4.034.

\bibitem{MB09}
G. Montani, R. Benini, \emph{Mod. Phys. L. A} \textbf{24}, 2667 (2009).

\bibitem{LM10}
M. Lattanzi, G. Montani, \emph{EuroPhys. L.} \textbf{89}, 39001 (2010).

\bibitem{LB96}
D. Lynden-Bell, \emph{Mon. Not. RAS} \textbf{279}, 389 (1996).

\bibitem{S08}
H.C. Spruit, \emph{Lect. Notes Phys.} \textbf{794}, 233 (2010).

\bibitem{SS01}
R. Stehle, H.C. Spruit, \emph{Mon. Not. RAS} \textbf{323}, 587 (2001).

\bibitem{F37}
V.C.A. Ferraro, \emph{Mon. Not. RAS} \textbf{97}, 458 (1937).

\end{thebibliography}
\end{document}